\documentclass[floatfix,twocolumn,aps,prl,showpacs]{revtex4}
\usepackage{natbib,graphicx,epsfig,amsfonts,bm,amsmath,amssymb}

\begin{document}
\title{Solving the Kadanoff-Baym equations for inhomogenous systems: 
Application to atoms and molecules}

\author{Nils Erik Dahlen}
\author{Robert van Leeuwen}
\affiliation{Theoretical Chemistry, Zernike Institute for Advanced Materials,
University of Groningen, Nijenborgh 4, 9747 AG Groningen, The
Netherlands}

\date{\today}

\begin{abstract}

We have implemented time-propagation of the non-equilibrium Green function 
for atoms and molecules, by solving the Kadanoff-Baym equations within a 
conserving self-energy approximation. 
We here demonstrate the usefulnes of time-propagation for calculating 
spectral functions and for describing
the correlated electron dynamics in a non-perturbative electric field.
We also demonstrate the use of time-propagation as a method for calculating
charge-neutral excitation energies,  equivalent to 
highly advanced solutions of the Bethe-Salpeter equation.

\end{abstract}

\maketitle

\newcommand{\be}{\begin{equation}}
\newcommand{\ee}{\end{equation}}
\newcommand{\bea}{\begin{eqnarray}}
\newcommand{\eea}{\end{eqnarray}}
\newcommand{\la}{\langle}
\newcommand{\ra}{\rangle}
\newcommand{\Tr}{\text{Tr}}
\newcommand{\br}{\mathbf{r}}
\newcommand{\bx}{\mathbf{x}}
\newcommand{\GKS}{G_{\text{KS}}}
\newcommand{\GHF}{G_{\text{HF}}}
\newcommand{\GLDA}{G_{\text{LDA}}}
\newcommand{\GOEP}{G_{\text{OEP}}}
\newcommand{\GKSI}{G^{-1}_{\text{KS}}}
\newcommand{\vks}{v_{\text{KS}}}
\newcommand{\vxc}{v_{\text{xc}}}
\newcommand{\half}{\frac{1}{2}}
\newcommand{\LW}{\text{LW}}
\newcommand{\HF}{\text{HF}}
\newcommand{\Chi}{\raisebox{.5ex}{\large{$\chi$}}}
\newcommand{\Ga}{G^k_{\text{HF}}}

The arrival of molecular electronics has exposed the need for improved methods 
for first-principles calculations on non-equilibrium quantum 
systems \cite{xue02}. The non-equilibrium Green function 
\cite{kadanoff62,keldysh65} is for several reasons
a natural device in such studies. Not only is it relatively simple,
being a function of two coordinates, but it contains a wealth of 
information, including the electron density and current, the total energy, 
ionization potentials, and excitation energies. Time-propagation 
according to the Kadanoff-Baym equations 
\cite{kadanoff62,schaefer96} is a direct method for describing the correlated electron
dynamics, and a method which automatically
leads to internally consistent and unambiguous results in agreement with 
macroscopic conservation laws. 
In the linear response regime, time-propagation within relatively
simple self-energy approximations corresponds to solving the Bethe-Salpeter
equation with highly advanced kernels \cite{kwong00}. 
The Green function techniques are also highly interesting as a 
complementary method to time-dependent density functional theory (TDDFT)
\cite{runge84,tddftbook}, not only by providing benchmark results for testing new TDDFT 
functionals, but diagrammatic techniques can also be used to systematically 
derive improved density functionals \cite{vanleeuwen96}. 
We will in this Letter demonstrate time-propagation for inhomogenous systems,
using the beryllium atom and the H$_2$ molecule as illustrative 
examples. 
 
The non-equilibrium Green function $G(\bx t, \bx' t')$ depends on 
two time-variables, rather than one such as  the 
time-dependent many-particle wave function. On the other hand, the fact that
it also depends on only two space and spin variables $\bx=(\br,\sigma)$ means
that it can be used for calculations on systems that are too large for solving
the time-dependent Schr\"odinger equation, such as the homogenous electron gas
\cite{kwong00} or solids\cite{haug98}, or for calculations on molecular conduction. In addition, the
Green function provides information about physical properties such as 
ionization potentials and spectral functions, that are not given by the 
many-particle wave function.
The Green function techniques also have  the advantage over, e. g., TDDFT or
density matrix methods based on the BBGKY hierarchy \cite{bonitz98}, that
it is easy to find approximations which give observables in agreement with
macroscopic conservation laws.

\begin{figure}
\begin{center}
\includegraphics[width=8.6cm]{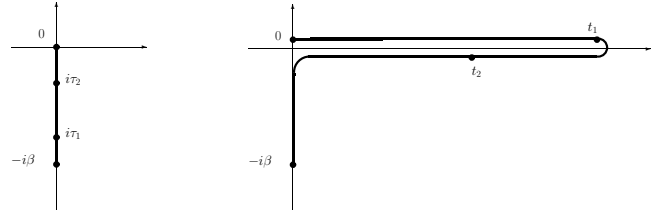}
\end{center}
\caption{At $t=0$, the system is in thermal equilibrium,
and the Green function is calculated for imaginary
times from $0$ to $-i\beta$ (left figure).
Describing the system for $t>0$ implies calculating
$G(t_1, t_2)$ on an extending time-contour (right).} \label{fig:contour}
\end{figure}

The two time-arguments of the Green function are located on a time-contour as
illustrated in Fig.~\ref{fig:contour}. It solves 
the equation of motion (we suppress the space- and spin-variables for 
notational simplicity)
\be
\left[i\partial_t -h(t)\right] G(t,t')= \delta(t,t') + \int_C d\bar t \,
\Sigma(t,\bar t) G(\bar t, t'),\label{eq:kbe}
\ee
where $h(t)$ is the non-interacting part of the hamiltonian (which may include
an arbitrarily strong time-dependent potential), and the self-energy $\Sigma$
accounts for the effects of the electron interaction. 
We use atomic units throughout. The time-integral is performed along the 
contour, and the delta-function $\delta(t,t')$ is defined on the contour 
\cite{danielewicz84}. We  only consider systems initially 
(at $t=0$) in the ground state. This is facilitated by describing the system
within the finite-temperature formalism 
\cite{note1},
letting the time-contour start at 
$t=0$ and end at the imaginary time $t=-i\beta=-i/(k_B T)$, as illustrated in
Fig.~\ref{fig:contour}. 
This choice of initial conditions means that the first step consists of 
calculating the Green function for both time-arguments
on the imaginary track.
The calculations are carried out in a basis of Hartree-Fock (HF) 
molecular orbitals, such that 
$G(\bx t, \bx't')=\sum_{i,j}\phi_i(\bx) G_{ij}(t,t') \phi_j^*(\bx')$. 
The Green function is consequently represented as a time-dependent matrix,
and the equations of motion reduce to a set of 
coupled matrix equations. The HF orbitals $\phi_i(\bx)$ are 
themselves given by linear combinations of Slater functions centered on 
the nuclei of the molecule.

\begin{figure}
\begin{center}
\includegraphics[width=8.6cm]{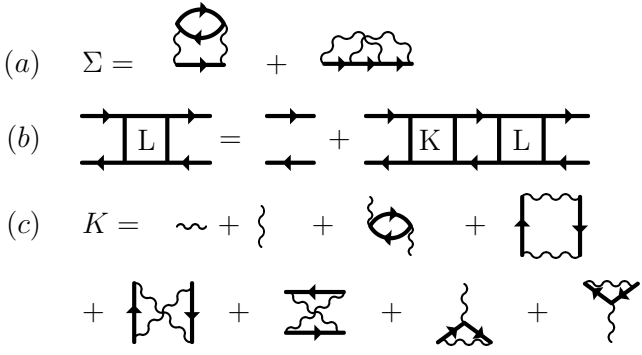}
\end{center}
\caption{The correlation part of the second-order self-energy (a), 
a diagrammatic representation of the Bethe-Salpeter equation (b), and the 
corresponding Bethe-Salpeter kernel (c).
The Green function lines represent self-consistent, full Green 
functions.} \label{fig:phi}
\end{figure}

We have solved the Kadanoff-Baym equations within the second-order 
self-energy approximation,
illustrated in Fig.~\ref{fig:phi}. This is a conserving 
approximation \cite{baym61,baym62}, which is essential for obtaining
results in agreement with macroscopic conservation 
laws. This is easily verified numerically, for instance by checking that the
total energy  remains constant when the Hamiltonian is 
time-independent.
For a given $G(t,t')$ matrix, the self-energy matrix
(for a spin-unpolarized system) is given by $\Sigma(t,t')=
\delta(t,t') \Sigma^{\text{HF}}(t)+\Sigma^{(2)}(t,t')$, where
\bea
\Sigma^{(2)}_{ij}(t,t')&=&\sum_{klmnpq} G_{kl}(t,t') G_{mn}(t,t') G_{pq}(t',t)
\nonumber\\
&&\times v_{iqmk} (2v_{lnpj}-v_{nlpj}),
\eea
and $\Sigma^{\text{HF}}_{ij}(t)=-i\sum_{kl} G_{kl}(t,t^+)
(2v_{ilkj}-v_{iljk})$ is the HF self-energy.
The two-electron integrals are defined by
\be
v_{ijkl}=\int \int d\bx  d\bx'\, 
\phi^*_i(\bx) \phi^*_j(\bx') v(\br-\br')\phi_k(\bx') \phi_l(\bx).
\ee
As the initial Hamiltonian 
is time-independent, the Green function on the imaginary track of the contour 
only 
depends on the difference between the two imaginary time-coordinates. Solving
for $G^M(i\tau-i\tau')\equiv G(-i\tau,-i\tau')$
(we use $\tau$ to denote time-arguments on the imaginary axis)
is then equivalent to 
solving the ordinary Dyson equation within the finite temperature formalism 
\cite{dahlen05b}. 
Note that the definition of $G^M$ in Ref.~\cite{dahlen06} differs from the one 
used here by a prefactor $i$.
Since the self-energy is a functional of the Green function, 
the Dyson equation should be solved to self-consistency.

Once the equilibrium Green function has been calculated, it can be
propagated according to the Kadanoff-Baym equations 
\cite{kadanoff62}. Time-propagation means that the contour which 
initially goes along the imaginary axis is extended 
along the real axis (see Fig.~\ref{fig:contour}).
The Green functions with both arguments on the real time-axis are denoted
by the functions 
$G_{ij}^>(t,t')
=-i\la \hat c_i(t) \hat c^\dagger_j(t') \ra$ 
(if $t$ is later on the contour than $t'$) and 
$G_{ij}^<(t,t')
=i\la \hat c^\dagger_j(t') \hat c_i(t) \ra$ 
(if $t'$ is later than $t$), with the symmetry 
$\left[ G^\lessgtr_{ij}(t,t')\right]^* =-G^\lessgtr_{ji}(t',t)$ and the 
boundary condition 
$G^>_{ij}(t,t)-G^<_{ij}(t,t)=-i\delta_{ij}$. We also need to calculate the 
functions $G^\rceil(t,i\tau)$ and $G^\lceil(i\tau,t)$ 
with one real and one imaginary time-argument.
The implementation of the propagation is similar to the scheme described
by K\"ohler \textit{et.  al.} in Ref.~\cite{koehler99}, with 
one important difference being that we are here dealing with
an inhomogenous system, i.e. the Green function, 
self-energy, and $h(t)$  are time-dependent matrices rather than vectors.

Another important difference with the propagation scheme described in 
Ref.~\cite{koehler99} is the initial correlations. In our 
case, they are given by the equilibrium
Green function according to
$G^<(0,0)=G^M(0^-)$ and
$G^\rceil(0,-i\tau)=G^M(-i\tau)$.
Due to the anti-periodicity 
 $G^M(i\tau+i\beta)=-G^M(i\tau)$, the resulting nonequilibrium Green function
will automatically satisfy the Kubo-Martin-Schwinger boundary condition
$G(0, t)=-G(-i\beta, t)$. 
The time-stepping follows the four coupled Kadanoff-Baym equations
\bea
\left[ i\partial_t - h(t)\right]  G^>(t, t') &=&
I^>_1(t,t')  ,\nonumber\\
\left[ i\partial_t - h(t)\right] G^\rceil(t,i\tau)
&=& I^\rceil(t,i\tau).
\eea
and the two corresponding adjoint equations \cite{dahlen06}.
The collision integrals are 
\bea
I^>_1(t,t')&=&
\int_0^t d\bar t \,
\Sigma^R(t,\bar t) G^>(\bar t, t') +  
\int_0^{t'} d\bar t \,
\Sigma^> (t, \bar t) G^A(\bar t, t') 
\nonumber\\
&&+ \frac{1}{i}\int_0^{\beta} d\bar \tau \, \Sigma^\rceil(t,-i\bar \tau) 
G^\lceil(-i\bar \tau, t') \label{eq:i1} \\
I^\rceil(t,i\tau) &=& \int_0^t d\bar t \,
\Sigma^R(t,\bar t) G^\rceil(\bar t, i \tau)  \nonumber\\
&& + \frac{1}{i} \int_0^{\beta} d\bar \tau \, 
\Sigma^\rceil (t, -i \bar \tau) G^M(i\bar \tau- i\tau)  \label{eq:i3}.
\eea
The retarded and advanced functions $G^{R/A}$ and $\Sigma^{R/A}$ are defined
according to $F^{R/A}(t,t')= \delta(t-t') F^\delta(t) 
\pm \theta(t-t') \left[F^>(t,t') -F^<(t,t') \right]$, where only the 
self-energy has a singular part (the HF self-energy). 
The last terms in each of Eqs.~(\ref{eq:i1}) and (\ref{eq:i3}) account for
the initial correlations of the system, and do not vanish when $t,t'\to0$.

We have been able to propagate the Green function for a number of closed-shell 
atoms and small diatomic molecules, where one can aim at 
quantitative agreement with experimental results. The kind of calculations
presented in this paper typically take 48 hours. 
The basis set must be large enough to describe the essential details of the 
electron dynamics. The 
most 
important limiting factor is the energy level structure of the systems; for
heavier atoms, the large eigenenergies of the core levels lead to rapid 
oscillations in the Green function, and one consequently needs to propagate
using time-steps much smaller than the time-scale of the interesting physical
phenomena dominated by the valence electrons. 
We have in these calculations used 28 basis 
functions for beryllium and 25 functions for H$_2$, with orbital 
energies lower than 5 Hartree. We include a time-dependent electric field in the 
direction of the molecular axis, so that the system 
preserves a cylindrical symmetry. Generalizing this
scheme to systems of lower symmetry does not lead to other complications than
increasing the size of the calculations. 
Since the systems 
considered here only have discrete energy levels, we do not observe strong 
damping effects such as what is observed in systems with a continuous spectrum 
\cite{koehler99,semkat99}. 

\begin{figure}
\includegraphics[width=8.6cm]{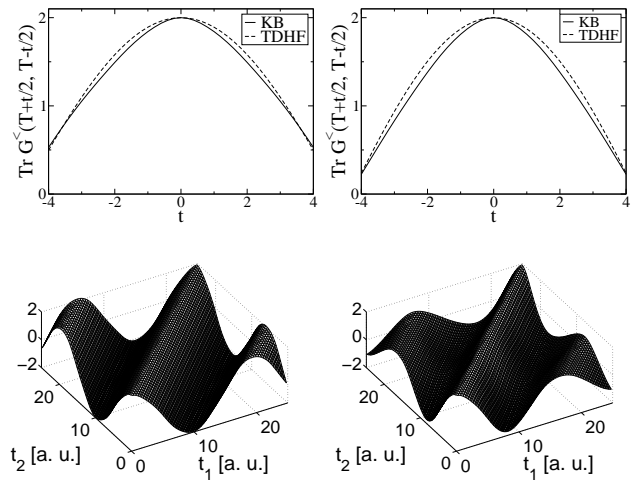}
\caption{The lower figures show the trace of the imaginary part of 
$G^<(t_1, t_2)$ for an H$_2$ molecule in its ground state (left) and in an
applied electric field (right). The Green function
in equilibrium only depends on $t_1-t_2$. The upper figures shows the same
quantity for a fixed value of $T=(t_1+t_2)/2$, compared with the same function
obtained from TDHF.}
\label{fig:h2A}
\end{figure}

Figure \ref{fig:h2A} shows the imaginary part of $\Tr \{G^<(t_1,t_2)\}$,
calculated for an H$_2$
molecule in equilibrium, and in the presence of a constant electric field 
$E(t)=\theta(t) E_0$ directed along the molecular axis. In both cases,
the trace of $G^<$ along the time-diagonal is constant, and equal to the 
particle number. In the ground-state (shown to the left), the Green function 
only depends on the difference 
$t_1-t_2$, and the oscillations perpendicular to the time-diagonal are given
by the ionization potentials of the molecule. The right figure illustrates
how the electric field ($E_0=0.14$ a.u.) changes the spectral properties of the
molecule. In addition to the expected narrowing of the ridge along the 
time-diagonal, the oscillations along the $t_1-t_2$ direction are damped. In
the upper figures, we show $\text{Im} \Tr \{ G^<(t_1, t_2)\}$ for a fixed 
$T=(t_1+t_2)/2=20$ a.u., compared with the same function calculated from 
time-dependent HF (TDHF). The ground state HF Green function has the form
$\Tr \{G_{\text{HF}}^<(t_1, t_2)\} =i \sum_i n_i e^{-i\epsilon_i(t_1-t_2)}$, and 
for the H$_2$ molecule we therefore have 
$\text{Im} \Tr \{G_{\text{HF}}^<(T+t/2,T-t/2)\}=2\cos(\epsilon_1 t)$. The correlated Green
function has a sharper peak at $t=0$ and while it is periodic in the relative
time-coordinate, it is characteristic of a correlated spectral function that it
can not be fitted to a cosine function at small $t$ \cite{bonitz98}. With
an added electric field, the energy fed into the system leads to a narrowing 
of the spectral function peak.

Time-propagation is also useful as a
direct method for calculating response functions and excitation 
energies \cite{kwong00}. The
excitation energies of the system can be obtained from the poles of the
density response function $\chi(\omega)$, defined by 
$\delta n(\omega)= \chi^R(\omega) \delta v(\omega)$.
Perturbing the system with a ``kick'' of the form $\delta v(t)=V \delta(t)$
excites all states compatible with the symmetry of the perturbing
potential $V$. For a kick in the form of an electric field along the molecular 
axis, the induced dipole moment is given by
$d(t)=E_0 \int d\br d\br' \, z \chi^R(\br,\br';t) z'$. The imaginary part of 
the Fourier transformed dipole moment then has peaks at the poles of 
$\chi^R(\omega)$, corresponding to the excitation energies of the system. 
Time-propagation is in this way an interesting, and far more direct alternative
to calculating the response function from $\chi(1,2)=L(1,2;1,2)$ where the 
particle-hole propagator $L$ is found by solving the Bethe-Salpeter equation 
\cite{baym61,baym62}, $ L=L_0+L_0 K L$, as illustrated diagrammatically in 
Fig.~\ref{fig:phi}.
The self-energy approximation used in this paper would correspond to the
kernel $K=\delta \Sigma/\delta G$ shown in Fig.~\ref{fig:phi}, where it should be noted that the Green
functions are the full
Green functions of the interacting system.
\begin{figure}
\begin{center}
\includegraphics[width=8.6cm]{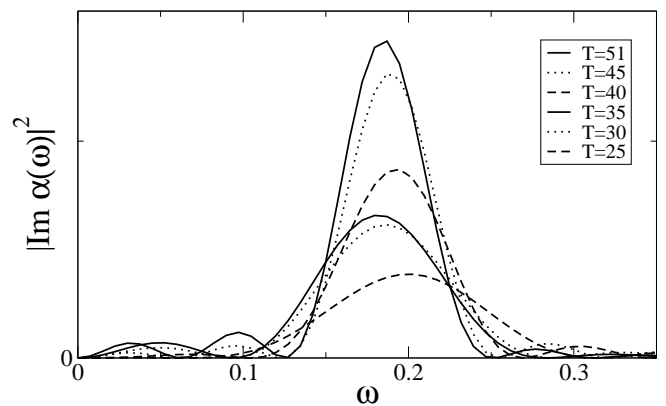}
\caption{The imaginary part of the polarizability $\alpha(\omega)$ of a 
beryllium atom calculated
from the Green function propagated up to a time $T$. } \label{fig:alpha}
\end{center}
\end{figure}
In Fig.~\ref{fig:alpha}, we have plotted the imaginary part of the 
polarizability, defined according to 
$\alpha_T(\omega)=-1/E_0 \int_0^T dt\, e^{i\omega t} d(t)$,
of a beryllium atom for various durations $T$ of the time-propagation. The 
polarizability develops a distinct peak at the $^1S$ $\to$ $^1P$  transition energy
$\omega = 0.189$ a.u. (compared to the TDHF value of 0.178
calculated within the same basis, and the experimental value of
0.194 \cite{kramida97}), which becomes increasingly sharp as the 
propagation time is extended. As the system consists only of discrete energy
levels, the damping of the time-dependent dipole moment $d(t)$ as a function of
time is not significant in the relatively short duration of the 
time-propagation. The position of the excitation energy
peak in Fig.~\ref{fig:alpha} therefore converges slowly, but extrapolation 
schemes nevertheless gives the position of the peak very accurately.

\begin{figure}
\begin{center}
\includegraphics[width=8.6cm]{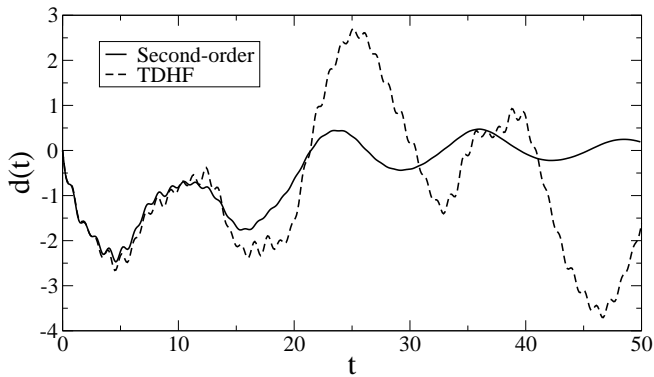}
\caption{The time-dependent dipole moment of a Be atom, calculated within the
HF and the second-order approximation.} \label{fig:alpha2}
\end{center}
\end{figure}

In Fig.~\ref{fig:alpha2}, we have plotted the time-dependent dipole moment of 
the beryllium atom, this time with a non-perturbative ``kick'' at $E_0=1.0$ 
a.u.. In this case, we see a clear difference between the uncorrelated 
Hartree-Fock result and the dipole moment calculated from the Kadanoff-Baym 
equations. 
The difference between the correlated and the uncorrelated results become
more apparent with time, due to the non-linear effects entering via the
self-energy.
Since we have now excited all unoccupied states, we can expect some damping
in the time-dependent dipole moment, but the damping is much more pronounced
in the correlated calculation.

In conclusion, we have demonstrated that time-propagation of the 
non-equilibrium Green function can be used as a practical method for 
calculating non-equilibrium and equilibrium properties of atoms and
molecules (e. g. for atoms in strong laser fields). 
For these small systems, the second order approximation
is clearly well-suited.
For extended systems, such as molecular
chains, it becomes essential to cut down the long range of the Coulomb
interaction. This is done effectively within the $GW$ approximation
\cite{hedin65,aryasetiawan98} (known as the shielded potential approximation in 
Refs.~\cite{baym61,baym62}) where the self-energy is given as a product of
the Green function $G$ and the dynamically screened interaction $W$ and which we
have currently implemented for the ground state~\cite{stan06}. For calculations involving
heavier atoms it will be necessary to use pseudopotentials in order to avoid
the very short time-steps necessary to account for the core-levels. For larger
molecules or solids, the calculations are certainly feasible in the framework
of model hamiltonians that include electron interaction. Green function
calculations are therefore highly important 1) for providing benchmark results 
for simpler methods, 2) for investigating the role of electron correlation, 
and 3) as and alternative to solving the Bethe-Salpeter equation with highly
sophisticated kernels.


\begin{thebibliography}{25}
\bibitem[{not({\natexlab{a}})}]{note1}
\bibinfo{note}{We here consider only finite systems, for which the
  finite-temperature formalism is used as a formal device for imposing boundary
  conditions on the Green function, but is only meaningful in the $T\to 0 $
  limit.}
\end{thebibliography}
\end{document}